\documentclass[manuscript]{aastex}

\usepackage{amsmath,amssymb}
\usepackage{graphicx}
\usepackage{graphics}

\shorttitle{Solar prominences}
\shortauthors{Keppens et al.}

\begin{document}

\title{Solar prominences: `double, double \ldots boil and bubble'}

\author{R. Keppens\altaffilmark{1} and C. Xia}
\affil{Centre for mathematical Plasma Astrophysics, Department of 
Mathematics, KU Leuven,\\
Celestijnenlaan 200B, 3001 Leuven, Belgium}
\email{Rony.Keppens@wis.kuleuven.be}

\and

\author{O. Porth}
\affil{Department of Applied Mathematics, The University of Leeds
, Leeds, LS2 9JT, UK}
\altaffiltext{1}{Concurrent Professor, School of Astronomy and Space Science, Nanjing University, Nanjing 210093, PR China}

\begin{abstract}
Observations revealed rich dynamics within prominences, the cool ($10^4$ K), macroscopic (sizes of order 100 Mm) `clouds' in the million degree solar corona. Even quiescent prominences are continuously perturbed by hot, rising bubbles. Since prominence matter is hundredfold denser than coronal plasma, this bubbling is related to Rayleigh-Taylor instabilities. Here we report on true macroscopic simulations well into this bubbling phase, adopting a magnetohydrodynamic description from chromospheric layers up to 30 Mm height. Our virtual prominences rapidly establish fully non-linear (magneto)convective motions where hot bubbles interplay with falling pillars, with dynamical details including upwelling pillars forming within bubbles. Our simulations show impacting Rayleigh-Taylor fingers reflecting on transition region plasma, ensuring that cool, dense chromospheric material gets mixed with prominence matter up to very large heights. This offers an explanation for the return mass cycle mystery for prominence material. Synthetic views at
extreme ultraviolet wavelengths show remarkable agreement with observations, with clear indications of shear-flow induced fragmentations.
\end{abstract}

\keywords{magnetohydrodynamics (MHD) --- Sun: filaments, prominences --- Sun: corona}

\section{Rayleigh-Taylor in prominences}\label{intro}
Prominences are among `the most common features of the solar atmosphere'~\citep{parenti14} and are a consequence of the Lorentz force levitating solar plasma against gravity. This creates density inversions in the hot solar corona favourable to Rayleigh-Taylor driven dynamics. Rayleigh-Taylor instability -- the reason why water falls out of a cup turned upside down -- can occur whenever a fluid or gas gets accelerated or pushed into a denser substance and is at the heart of many dynamical phenomena in astrophysical plasmas. Recent magnetohydrodynamic (MHD) modeling of Rayleigh-Taylor evolution for the Crab nebula showed a clear tendency to self-organize into larger-scale structures, with filament sizes reaching up to a quarter of the entire pulsar wind nebula radius~\citep{porthrtcrab14}. Magnetic field-guided accretion processes onto magnetized stars are enriched by equatorially accreting Rayleigh-Taylor plasma tongues, protruding into the magnetosphere from the inner accretion disk edge~\citep{kulkarni08}. In the solar context, Rayleigh-Taylor filamentary structure can form during flux emergence and its reconnection with pre-existing coronal fields, as demonstrated by means of high resolution MHD simulations~\citep{isobe05}.

Solar prominences also demonstrate Rayleigh-Taylor mediated dynamics, with Hinode Solar Optical Telescope observations~\citep{berger08} revealing how quiescent prominences show bright downflowing filaments of several hundred kilometres in width, typical speeds of ${\cal O}(10) \,\mathrm{km}/\mathrm{s}$ and ten minute lifetimes. At the same time, dark inclusions mark upflows at 20 $\mathrm{km}/\mathrm{s}$, rising up to 18 Mm heights and shedding voids that in turn grow to Mm sizes. Detailed observations showed how such dark upflows originate at the top of the chromosphere and can grow to 4-6 Mm plumes and rise to 15 Mm heights~\citep{berger10}. They can form large-scale (20-50 Mm) buoyant arches or bubbles, and these rising bubbles were found to contain 25-120 times hotter material than the prominence proper~\citep{berger11}, strengthening the argument for a magneto-thermal convection process typical for coronal cavity-prominence configurations. Using a local model for a dipped prominence bottom boundary, Rayleigh-Taylor mode development in three-dimensional (3D) MHD simulations demonstrated both upflows~\citep{hillier12} and interchange reconnection leading to downward blob motions~\citep{hillier12b}, in general agreement with observed local details. Recent modeling efforts have included partial ionization effects in local box models of the prominence-corona transition region~\citep{khomenko14}, finding clear differences with pure MHD approaches, as neutrals experience faster descents. Recent theoretical findings quantified the potentially stabilizing role of magnetic shear in idealized incompressible settings, important in the linear stages of Rayleigh-Taylor activity~\citep{ruderman14}. A step towards global modeling of prominence dynamics in arcade systems confirmed this role of sheared magnetic fields as well as the effects of line-tying on prominence stability~\citep{terradas15}, but lacked the resolution to follow their development into the strongly nonlinear regime. In this paper, we set forth to realize this step, for the first time including chromosphere-transition region variations in  a 3D prominence setup, and simulating well into the observed magneto-thermal convective motions.

\section{Numerical setup}\label{setup}
Our simulation box extends for 30 Mm horizontally ($x$) and vertically ($y$), and has a width ($z$) of 14 Mm. Using three levels of dynamic grid refinement we achieve a resolution of $600\times 600\times 280$, i.e. grid cell sizes down to 50 km. With the open-source MPI-AMRVAC software~\citep{amrvac12,amrvac14}, we perform 3D, ideal MHD simulations for two cases that differ most markedly in the prevailing magnetic field strength throughout the domain: a weak field case at $\approx 8$ G, and a stronger field one at $\approx 20$ G. These are representative values for quiescent prominence conditions. 
The initial magnetic field $\mathbf{B}=(B_x=0.1998 \,\mathrm{G},0,B_z(y))$  is purely planar and non-uniform, due to an exponential decrease of the strongest $B_z(y)$ component in a layer of 2.5 Mm thickness above the initial prominence heigth $y_\mathrm{p}=12.5 \,\mathrm{Mm}$. The analytic form for the initial $B_z(y)$ is given by
\begin{eqnarray}
B_z & = & B_{z0} \,\,\,\,\,\,\, \mathrm{for} \,\, y<y_\mathrm{p} \,, \nonumber \\
 & = & B_{z0} \exp\left[-(y-y_\mathrm{p})/\lambda_{B}\right]\,\,\,\,\, \mathrm{for}\,\, y_\mathrm{p}\leq y \leq y_\mathrm{b} \,\, \nonumber \\
 & = & B_{z0} \exp\left[-(y_\mathrm{b}-y_\mathrm{p})/\lambda_{B}\right]\,\,\,\,\, \mathrm{for}\,\, y_\mathrm{b}\leq y  \,. 
\end{eqnarray}
The parameters are set to $\lambda_B=15$ Mm, and $y_\mathrm{b}=15$ Mm, while $B_{z0}$ differs between the weak and strong field case.
This induces a local shear in magnetic field, and establishes an upward magnetic pressure force that lifts matter against solar gravity. The horizontal $B_x$ field component leads to stabilization by tension forces against purely planar $(x,y)$ Rayleigh-Taylor instabilities for all wavelengths exceeding about 33 km, slightly below our numerical resolution. Nonlinear effects quickly dominate the dynamics at lengthscales fully captured in our study, a result corroborated by high resolution purely two-and-a-half dimensional simulations. The initial magneto-hydrostatic stratification introduces a transition region height at $y_{\mathrm{tr}}=2.5 \,\mathrm{Mm}$ where the temperature smoothly connects an $8000 \,\mathrm{K}$ chromosphere to a $1.8 \,\mathrm{MK}$ corona. 

Figure~1 shows in top and bottom left panels the temperature and density structure for the strongest magnetic field case. The dashed profiles above $y_\mathrm{p}=12.5 \,\mathrm{Mm}$ quantify $T(y)$ and $\rho(y)$ exterior to the prominence, where the corona is isothermal but the density shows an increase due to magnetic levitation. Inside the prominence,  the vertical stratification follows the solid curves shown in Figure~1: the prominence temperature is 12000 K in $y\in[y_\mathrm{p}, 15 \,\mathrm{Mm}]$, increases linearly with height to 60000 K at $25\, \mathrm{Mm}$, and there connects again to coronal temperatures above the prominence structure. This initial condition has the essential characteristics of solar filaments, as the density contrast $\rho_{\mathrm{prom}}/\rho_{\mathrm{cor}}$ below the prominence is about 127.3 in this strong field case. 
When we set the overall dimensions of the prominence segment at 30 Mm length and 5 Mm width, we find that 
the total prominence mass is $7.5 \times 10^{13} \mathrm{g}$ for the weak field case, going up to $2.9 \times 10^{14}\mathrm{g}$ for the strong field case. These masses, together with the overall dimensions, all fall within their observationally known ranges. 

\section{Global evolution}\label{evolve}
The initial condition -- though vertically in force balance -- is out of pressure equilibrium in the $z$-direction across the prominence structure. This leads to a transient phase of successive compressions of the prominence matter (and in the coronal region above it), with shock waves traversing the periodic $z$-direction. These alter the detailed temperature-density variations throughout prominence and coronal regions upward from $y=y_\mathrm{p}$, but largely retain their essential characteristics, keeping the total prominence mass and typical corona-prominence density and temperature contrasts. More importantly for our study targeting long-term prominence internal dynamics, these transverse motions quickly become dominated by vertical and horizontal ($x$) velocity components, as demonstrated in the right panel of Figure~1, where the (scaled) kinetic energy evolution is plotted for each velocity component, for the strong field case (the weak field case behaves similarly in its energetic evolution). After about 4 minutes, vertical motions (solid line) dominate in kinetic energy, and they saturate before 10 minutes. Lateral movements ($x$-direction, dotted line) peak at about 11 minutes, while we ran our models for close to 15 minutes. The growth in vertical kinetic energy directly relates to Rayleigh-Taylor modes throughout the prominence segment, which are triggered by a superposition of 50 small-amplitude velocity perturbations that fit the periodic $x$-direction with random phases. 
Each individual flow perturbation represents a planar $(v_x,v_y,0)$ incompressible velocity field, and is localized about the bottom prominence position $y_\mathrm{p}$ and its midplane $z=0$.

Figure~2 gives a clear overview of the resulting prominence deformation and dynamics, by collecting a number of depth-integrated views taken at 6.9 minutes. This figure is for the strong field case. Panels (a) and (c) provide views on the prominence when integrated along its length, showing its entire embedding within coronal plasma, while the nonlinear Rayleigh-Taylor development has created downward falling pillars that just reached transition region heights. Panel (a)  integrates an additionally advected scalar, where green values correspond to prominence matter, dark purple indicates chromosphere plasma, and white is used for coronal material. Panel (c) relates to the instantaneous temperature structure, with white indicating cool (chromospheric and prominence) matter, and red is used for coronal values. This panel also shows a thin layer of hot matter just above the prominence structure, which locates shock-heated, initially evacuated matter found there. Animated views for the entire simulation in the format of Figure~2 are provided as online material, where the mentioned transverse compressions and their transient nature become evident. From our earlier simulations of actual prominence formation due to chromospheric evaporation, thermal instability and runaway catastrophic cooling events~\citep{xia11,xia12,fang13,keppens14}, these transient shock waves mimic the rebound shock fronts found to result from siphon flow driven impacts on the prominence-corona transition region. These rebound shocks ultimately repeatedly impact on the prominence structure, as a result of successive reflections when they reach chromosphere-corona transition regions along the fieldlines. 

The edge-on views shown in panels (b) and (d) of Figure~2 contain the $z$-integrated density structure, clearly dominated by falling Rayleigh-Taylor pillars with widths of up to 1000 km, and bubbles of upwardly curved prominence segments with lateral dimensions between 3000-4000 km. The resulting magnetic field deformation is visualized in panel (b), where streamlines, colored by the tracer from panel (a), are given for the $z$-integrated in-plane magnetic field components. This shows how the falling pillars indeed interchange magnetic field structure, where we note that the prevailing plasma beta is typically 0.16 (for the strong field case, and 0.38 for the weak one). The main displacements, as also seen in Figure~1, rapidly turn vertical and lateral, in accord with interchange modes that try to minimize field line bending, as the dominant magnetic field component is $B_z$ at all times. The same information can be deduced from panel (d) in Figure~2, where the superposed arrows likewise quantify the ($z$-integrated in-plane) velocity structure. Clearly, regions with Rayleigh-Taylor fingers are overall downward-moving at this time, while hot coronal plasma shows the fastest upwelling flows within several of the bubbles. An interesting detail is the upwelling Rayleigh-Taylor finger seen in panels (b)-(d) at horizontal distance $x\approx 12.5\,\mathrm{Mm}$ that is seen to start at $y\approx 10 \,\mathrm{Mm}$ working its way upwards from the bottom region of a bubble. This bubble has just been closed from below, by the merging of two downwelling fingers that jointly continue their fall. A localized dense protrusion then swirls upwards, indicating that the relative acceleration (between light and dense matter) causing the Rayleigh-Taylor event now acts upwards in the bottom region of this bubble. A three-dimensional view on the prominence structure is given for about the same time in the left panel of Figure~3. This shows the temperature variation in vertical bounding planes at $x=0$ and $z=-7 \,\mathrm{Mm}$. It also shows an isosurface of the temperature marking the 30000 K isosurface, showing that all cool material is found in the downward pillars and at the lower regions of the bubbles. This isosurface also nicely traces out the location of the chromosphere-corona transition region, which has hardly been perturbed at this point in the evolution. The grey isosurface shows the rear-half of the tracer isosurface, at a value which locates the original prominence matter at all times, as well as the chromosphere-corona transition. In this view, we also see several of the upwelling Rayleigh-Taylor features, in the closed bubble discussed earlier but also near the $x\approx 30\,\mathrm{Mm}$ front end. Hence such temporary upwelling features with widths of about 500 km, should be identifiable in the early stages of prominence formation and their internal dynamics. Note that speeds associated with individual larger-scale features, such as the falling and rising filaments or bubbles, are several tens of $\mathrm{km}/\mathrm{s}$, also seen from the animated views provided. Using the tracer mentioned earlier, we quantify a prominence-material-only average vertical speed. This increases from zero up to about $-20\,\mathrm{km}/\mathrm{s}$ after 400 seconds, declining again afterwards. To quantify better actual speeds, we added to our 3D MHD simulation a set of $30\times 60\times 10$ Lagrangian particles, which originally are positioned regularly on a grid throughout the simulation box. In the snapshot shown in Figure~3, twenty-four of the 18000 particle trajectories obtained are visualized with streamlines that color from dark to white when time proceeds through the almost 15 minute interval simulated. These select initial locations all near the midplane $z=0$, half of them initially right below the $y=y_\mathrm{p}$ bottom prominence layer, while the other 12 are internal to the prominence. The right panel of Figure~3 shows the same set of particle trajectories, but viewed  in $(v_y,y)$ phase space. Dotted horizontal lines mark the heights of the initial prominence edge $y_p$, as well as the transition region height at 2.5 Mm. In this view, the ones that started internal to the prominence are indicated with larger symbols, while the thinner trajectories correspond to external (coronal) matter. Figure~3 shows that downward motions at up to 60 $\mathrm{km}/\mathrm{s}$ prevail at first, but nearly all get deflected upwards after encountering the transition region. Both coronal and internal prominence matter can get accelerated up to speeds exceeding 120 $\mathrm{km}/\mathrm{s}$. They can thereby reach heights well above their starting position, as some tracks go beyond 20 Mm height. Since a typical sound speed for the corona is 200 $\mathrm{km}/\mathrm{s}$, while the internal prominence sound speed is ten times lower, the process is highly nonlinear, and a vigorous magnetoconvective flow pattern extends from the transition region up into the prominence surroundings. Prominence matter can thereby repeatedly recycle, as it traverses a large range in altitude. These Lagrangian trajectories also imply that field lines (mainly directed along $z$) indeed show significant interchange behavior. This aspect may be exaggerated in our simulation by the periodicity assumption: in reality, these field lines are part of an arcade system passing through the prominence matter, and line-tying effects play a role in determining their Rayleigh-Taylor stability properties~\citep{terradas15}. 

An impression of the magnetoconvection that gets established throughout the prominence surroundings is shown in Figure~4, where the high field case is visualized at the endtime of our simulation, i.e. at 14.3 minutes. At left, we show the tracer isosurfaces that identify all prominence matter (colored by the local temperature), along with a grey isosurface that identifies the original chromosphere-corona transition region. The former isosurface shows that prominence bubbles have merged and grown into arch-shaped structures that can reach sizes up to 10 Mm in width. The latter isosurface demonstrates that the impacting Rayleigh-Taylor fingers can locally significantly perturb the transition region, and cause dense chromospheric matter to be hurled up to heights of 10 Mm or more. This provides an effective route to feed more cool, dense matter into the prominence environment, and hence plays an important role in its mass recycling. Figure~4 also shows the density structure in a vertical slicing plane at $z=0$ in the box at right. In this view, we also visualize all tracer particles found between $x=15$ and 30 Mm initially, where their color encodes the original starting height of the particle. At time zero, this color coding gives a plane-parallel green-orange-yellow-red distribution from top to bottom, while at the endtime from Figure~4, vigorous convection shows effective mixing in the entire region between 2.5 Mm and up to 23 Mm. While Figure~4 is for the high field strength case, we provide animated views for the low field strength case in the representation of Figure~4, as online material. Qualitatively similar trends occur in high and low field strength cases, although the maximal velocities attained are lower for the low field case, and the falling pillars reach the transition region a bit later in the evolution. This dynamical evolution allows one to interpret the temporal evolution of the component-wise kinetic energy through the box shown in Figure~1. The maximum correlates with impacting falling pillars on the transition region, and lateral deflections maximize when up and down welling material meet up.

\section{Synthetic observations and conclusions}\label{conc}
Our macroscopic simulations can be turned into extreme ultraviolet synthetic images, for direct comparison with those available from Solar 
Dynamics Observatory~\citep{sdo12} (SDO) observations using the Atmospheric 
Imaging Assembly~\citep{aia12} (AIA). Especially its 304 \AA\, and 171 \AA\, channels provide views highlighting matter at 80000 K and  800000 K, respectively. This then samples cooler prominence to transition region material. A synthetic view of both the high field (top rows) and low field case (bottom rows) in both EUV channels is given in Figure~5 at the endtime of our simulations, while animated views on the final seven minutes of evolution are provided online. One notices how cool prominence matter is found embedded in hotter material, with falling and rising structures over a fair range of lengthscales. The different wavelength channels show morphological differences between hot and cool, up and downflow streams. The simulated, late nonlinear stages, especially for the low field case, show clear substructure developing along the edges of the largest bubbles, as seen in the bottom panels of Figure~5. At this stage, strong shear flows are established all along the arcs, that now extend as 10 Mm wide structures to heights of 18 Mm. We expect similar details to develop in the later stages of the high field strength case as well, as it also shows strong shear flows. This is in direct agreement with the latest observational details, pointing to Kelvin-Helmholtz and Rayleigh-Taylor interplay at the bubble boundary~\citep{bergerproc14}. Visualizations of also the coronal channels (193 \AA\, and 211 \AA) further reveal the complex multi-temperature structure found in the magnetoconvective dynamics. Note that, by construction, our side-on views show the prominence matter in emission, and assume that the radiation is optically thin. This, together with the pure ideal MHD nature of our simulations, thereby neglecting important effects like coronal radiative losses, is an aspect to be improved upon. Further work  can strive for even higher resolutions to capture smaller-scale fine-structure development from interplaying shear flow-driven, gravitational and thermal instabilities, or modifications due to partial ionization conditions. Ultimately, ab initio simulations must be able to demonstrate the thermal instability mediated formation process of prominences~\citep{xialetter14}, and simultaneously capture Rayleigh-Taylor mode development in realistic fluxrope-embedded prominence structures.

\acknowledgments
This research was supported by the Interuniversity Attraction Poles
Programme (initiated by the Belgian Science Policy Office, IAP P7/08 CHARM) and by the KU Leuven GOA/2015-014. Simulations used the VSC (Flemish Supercomputer Center) funded by the
Hercules foundation and the Flemish government, and PRACE resources on SuperMUC at Garching. C.X. is supported by FWO Pegasus financing.

\bibliographystyle{apj}

\begin{thebibliography}{22}
\expandafter\ifx\csname natexlab\endcsname\relax\def\natexlab#1{#1}\fi

\bibitem[{{Berger}(2014)}]{bergerproc14}
{Berger}, T. 2014, in IAU Symposium, Vol. 300, IAU Symposium, ed.
  B.~{Schmieder}, J.-M. {Malherbe}, \& S.~T. {Wu}, 15--29

\bibitem[{{Berger} {et~al.}(2011){Berger}, {Testa}, {Hillier}, {Boerner},
  {Low}, {Shibata}, {Schrijver}, {Tarbell}, \& {Title}}]{berger11}
{Berger}, T., {Testa}, P., {Hillier}, A., {Boerner}, P., {Low}, B.~C.,
  {Shibata}, K., {Schrijver}, C., {Tarbell}, T., \& {Title}, A. 2011, Nature,
  472, 197

\bibitem[{{Berger} {et~al.}(2008){Berger}, {Shine}, {Slater}, {Tarbell},
  {Title}, {Okamoto}, {Ichimoto}, {Katsukawa}, {Suematsu}, {Tsuneta}, {Lites},
  \& {Shimizu}}]{berger08}
{Berger}, T.~E., {Shine}, R.~A., {Slater}, G.~L., {Tarbell}, T.~D., {Title},
  A.~M., {Okamoto}, T.~J., {Ichimoto}, K., {Katsukawa}, Y., {Suematsu}, Y.,
  {Tsuneta}, S., {Lites}, B.~W., \& {Shimizu}, T. 2008, Astrophys. J. Lett.,
  676, L89

\bibitem[{{Berger} {et~al.}(2010){Berger}, {Slater}, {Hurlburt}, {Shine},
  {Tarbell}, {Title}, {Lites}, {Okamoto}, {Ichimoto}, {Katsukawa}, {Magara},
  {Suematsu}, \& {Shimizu}}]{berger10}
{Berger}, T.~E., {Slater}, G., {Hurlburt}, N., {Shine}, R., {Tarbell}, T.,
  {Title}, A., {Lites}, B.~W., {Okamoto}, T.~J., {Ichimoto}, K., {Katsukawa},
  Y., {Magara}, T., {Suematsu}, Y., \& {Shimizu}, T. 2010, Astrophys. J., 716,
  1288

\bibitem[{{Fang} {et~al.}(2013){Fang}, {Xia}, \& {Keppens}}]{fang13}
{Fang}, X., {Xia}, C., \& {Keppens}, R. 2013, Astrophys. J. Lett., 771, L29

\bibitem[{{Hillier} {et~al.}(2012{\natexlab{a}}){Hillier}, {Berger}, {Isobe},
  \& {Shibata}}]{hillier12}
{Hillier}, A., {Berger}, T., {Isobe}, H., \& {Shibata}, K. 2012{\natexlab{a}},
  Astrophys. J., 746, 120

\bibitem[{{Hillier} {et~al.}(2012{\natexlab{b}}){Hillier}, {Isobe}, {Shibata},
  \& {Berger}}]{hillier12b}
{Hillier}, A., {Isobe}, H., {Shibata}, K., \& {Berger}, T. 2012{\natexlab{b}},
  Astrophys. J., 756, 110

\bibitem[{{Isobe} {et~al.}(2005){Isobe}, {Miyagoshi}, {Shibata}, \&
  {Yokoyama}}]{isobe05}
{Isobe}, H., {Miyagoshi}, T., {Shibata}, K., \& {Yokoyama}, T. 2005, Nature,
  434, 478

\bibitem[{{Keppens} {et~al.}(2012){Keppens}, {Meliani}, {van Marle}, {Delmont},
  {Vlasis}, \& {van der Holst}}]{amrvac12}
{Keppens}, R., {Meliani}, Z., {van Marle}, A.~J., {Delmont}, P., {Vlasis}, A.,
  \& {van der Holst}, B. 2012, Journal of Computational Physics, 231, 718

\bibitem[{{Keppens} \& {Xia}(2014)}]{keppens14}
{Keppens}, R. \& {Xia}, C. 2014, Astrophys. J., 789, 22

\bibitem[{{Khomenko} {et~al.}(2014){Khomenko}, {D{\'{\i}}az}, {de Vicente},
  {Collados}, \& {Luna}}]{khomenko14}
{Khomenko}, E., {D{\'{\i}}az}, A., {de Vicente}, A., {Collados}, M., \& {Luna},
  M. 2014, Astron. \& Astroph., 565, A45

\bibitem[{{Kulkarni} \& {Romanova}(2008)}]{kulkarni08}
{Kulkarni}, A.~K. \& {Romanova}, M.~M. 2008, Monthly Not. Roy. Astron. Soc.,
  386, 673

\bibitem[{{Lemen} {et~al.}(2012){Lemen}, {Title}, {Akin}, {Boerner}, {Chou},
  {Drake}, {Duncan}, {Edwards}, {Friedlaender}, {Heyman}, {Hurlburt}, {Katz},
  {Kushner}, {Levay}, {Lindgren}, {Mathur}, {McFeaters}, {Mitchell}, {Rehse},
  {Schrijver}, {Springer}, {Stern}, {Tarbell}, {Wuelser}, {Wolfson}, {Yanari},
  {Bookbinder}, {Cheimets}, {Caldwell}, {Deluca}, {Gates}, {Golub}, {Park},
  {Podgorski}, {Bush}, {Scherrer}, {Gummin}, {Smith}, {Auker}, {Jerram},
  {Pool}, {Soufli}, {Windt}, {Beardsley}, {Clapp}, {Lang}, \&
  {Waltham}}]{aia12}
{Lemen}, J.~R., {Title}, A.~M., {Akin}, D.~J., {Boerner}, P.~F., {Chou}, C.,
  {Drake}, J.~F., {Duncan}, D.~W., {Edwards}, C.~G., {Friedlaender}, F.~M.,
  {Heyman}, G.~F., {Hurlburt}, N.~E., {Katz}, N.~L., {Kushner}, G.~D., {Levay},
  M., {Lindgren}, R.~W., {Mathur}, D.~P., {McFeaters}, E.~L., {Mitchell}, S.,
  {Rehse}, R.~A., {Schrijver}, C.~J., {Springer}, L.~A., {Stern}, R.~A.,
  {Tarbell}, T.~D., {Wuelser}, J.-P., {Wolfson}, C.~J., {Yanari}, C.,
  {Bookbinder}, J.~A., {Cheimets}, P.~N., {Caldwell}, D., {Deluca}, E.~E.,
  {Gates}, R., {Golub}, L., {Park}, S., {Podgorski}, W.~A., {Bush}, R.~I.,
  {Scherrer}, P.~H., {Gummin}, M.~A., {Smith}, P., {Auker}, G., {Jerram}, P.,
  {Pool}, P., {Soufli}, R., {Windt}, D.~L., {Beardsley}, S., {Clapp}, M.,
  {Lang}, J., \& {Waltham}, N. 2012, Solar Phys., 275, 17

\bibitem[{{Parenti}(2014)}]{parenti14}
{Parenti}, S. 2014, Living Reviews in Solar Physics, 11, 1

\bibitem[{{Pesnell} {et~al.}(2012){Pesnell}, {Thompson}, \&
  {Chamberlin}}]{sdo12}
{Pesnell}, W.~D., {Thompson}, B.~J., \& {Chamberlin}, P.~C. 2012, Solar Phys.,
  275, 3

\bibitem[{{Porth} {et~al.}(2014{\natexlab{a}}){Porth}, {Komissarov}, \&
  {Keppens}}]{porthrtcrab14}
{Porth}, O., {Komissarov}, S.~S., \& {Keppens}, R. 2014{\natexlab{a}}, Monthly
  Not. Roy. Astron. Soc., 443, 547

\bibitem[{{Porth} {et~al.}(2014{\natexlab{b}}){Porth}, {Xia}, {Hendrix},
  {Moschou}, \& {Keppens}}]{amrvac14}
{Porth}, O., {Xia}, C., {Hendrix}, T., {Moschou}, S.~P., \& {Keppens}, R.
  2014{\natexlab{b}}, Astrophys. J. Suppl. Ser., 214, 4

\bibitem[{{Ruderman} {et~al.}(2014){Ruderman}, {Terradas}, \&
  {Ballester}}]{ruderman14}
{Ruderman}, M.~S., {Terradas}, J., \& {Ballester}, J.~L. 2014, Astrophys. J.,
  785, 110

\bibitem[{{Terradas} {et~al.}(2015){Terradas}, {Soler}, {Luna}, {Oliver}, \&
  {Ballester}}]{terradas15}
{Terradas}, J., {Soler}, R., {Luna}, M., {Oliver}, R., \& {Ballester}, J.~L.
  2015, Astrophys. J., 799, 94

\bibitem[{{Xia} {et~al.}(2012){Xia}, {Chen}, \& {Keppens}}]{xia12}
{Xia}, C., {Chen}, P.~F., \& {Keppens}, R. 2012, Astrophys. J., 748, L26

\bibitem[{{Xia} {et~al.}(2011){Xia}, {Chen}, {Keppens}, \& {van Marle}}]{xia11}
{Xia}, C., {Chen}, P.~F., {Keppens}, R., \& {van Marle}, A.~J. 2011, Astrophys.
  J., 737, 27

\bibitem[{{Xia} {et~al.}(2014){Xia}, {Keppens}, {Antolin}, \&
  {Porth}}]{xialetter14}
{Xia}, C., {Keppens}, R., {Antolin}, P., \& {Porth}, O. 2014, Astrophys. J.
  Lett., 792, L38

\end{thebibliography}

\clearpage

\begin{figure}
\includegraphics[width=\textwidth]{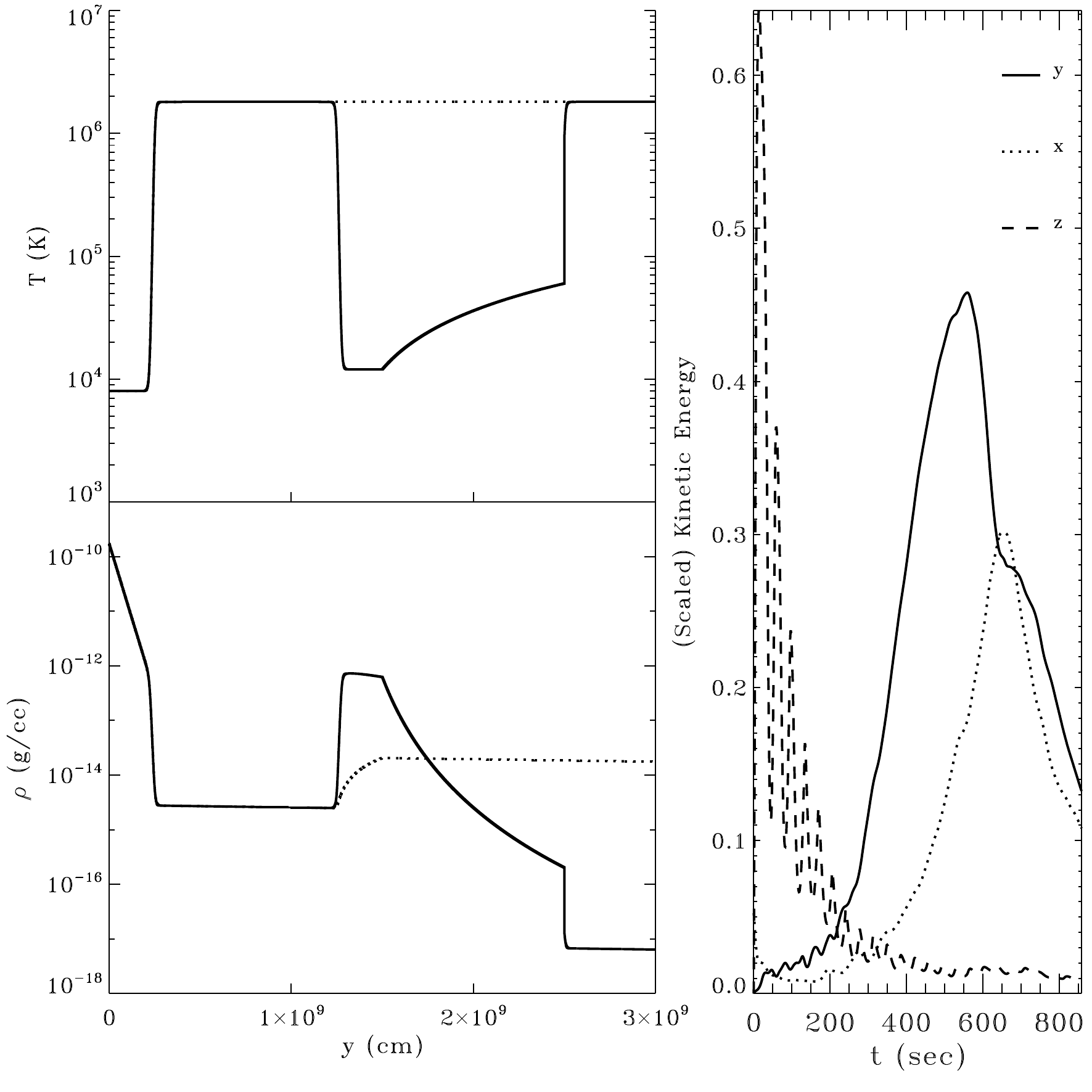}
\caption{The initial temperature (top left) and density (bottom left) stratification, both within (solid) and external (dotted) to the prominence. Right panel: the scaled kinetic energy evolution, plotted per velocity component: vertical (solid, $y$), lateral (dotted, $x$), and transverse (dashed, $z$).
}
\label{ffig1}
\end{figure}

\begin{figure}
\includegraphics[width=\textwidth]{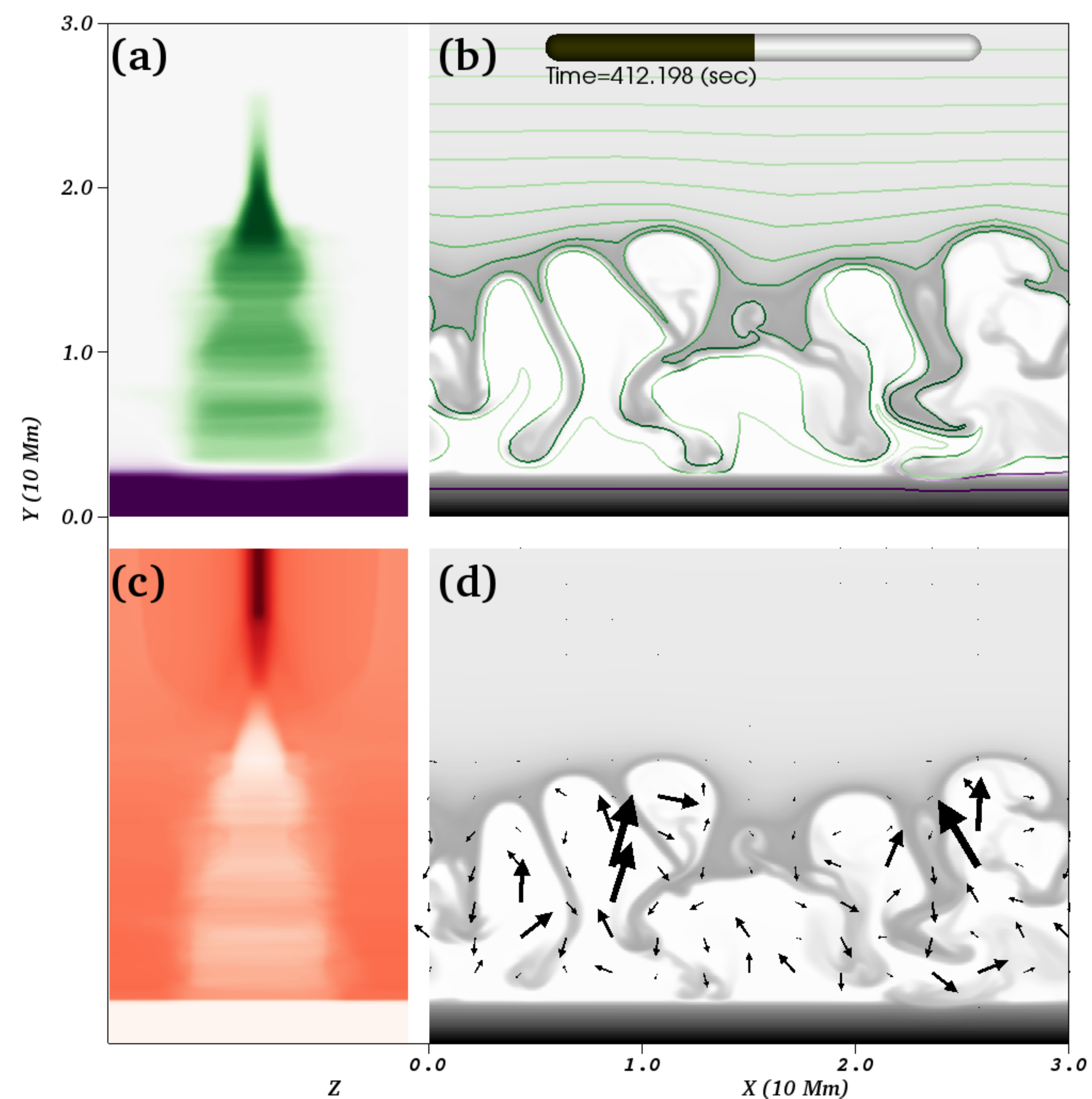}
\caption{Ray-traced views at about 6.9 minutes, along [panels (a), (c)] and across [panels (b), (d)] the prominence axis, showing contour views of: (a) the tracers used to identify prominence (green) and chromosphere (purple) material; (b)-(d) the density variation; (c) the temperature. Right top panel also shows fieldlines based on integrated horizontal magnetic components, while bottom right arrows quantify the in-plane velocity variation. A movie in this format (for the high field case) is provided online.}
\label{ffig2}
\end{figure}

\begin{figure}
\includegraphics[width=\textwidth]{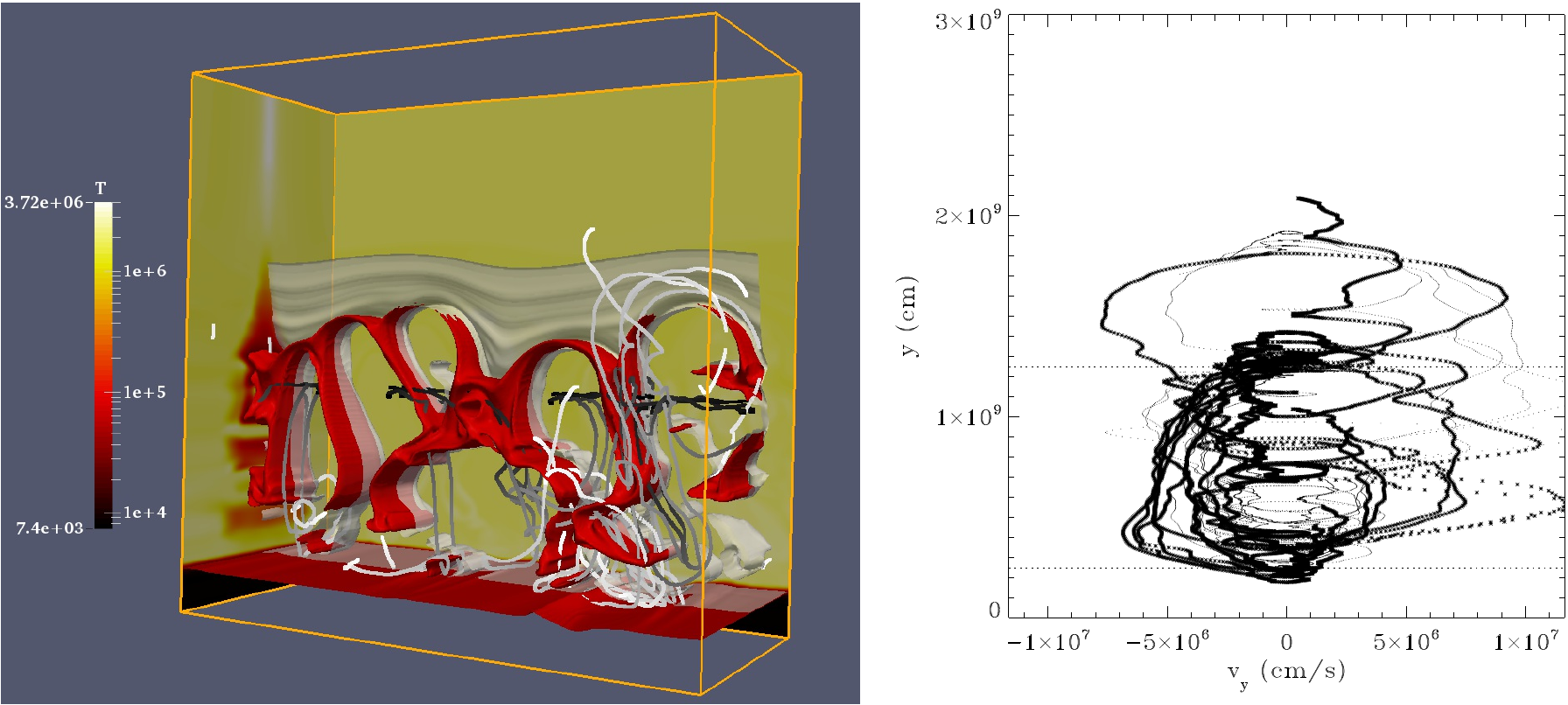}
\caption{Left: a 3D view on the prominence, at the same time as Figure~2, showing the temperature variation on bounding planes, as well as a (red) isocontour at 30000 K. The grey isosurface shows half of the prominence-bounding surface. Furthermore, 24 Lagrangian tracer paths are superposed, changing their color from black to white to indicate temporal variation. At right, the same 24 trajectories are displayed in a $(v_y,y)$ phase-space view.
}
\label{ffig3}
\end{figure}

\begin{figure}
\includegraphics[width=\textwidth]{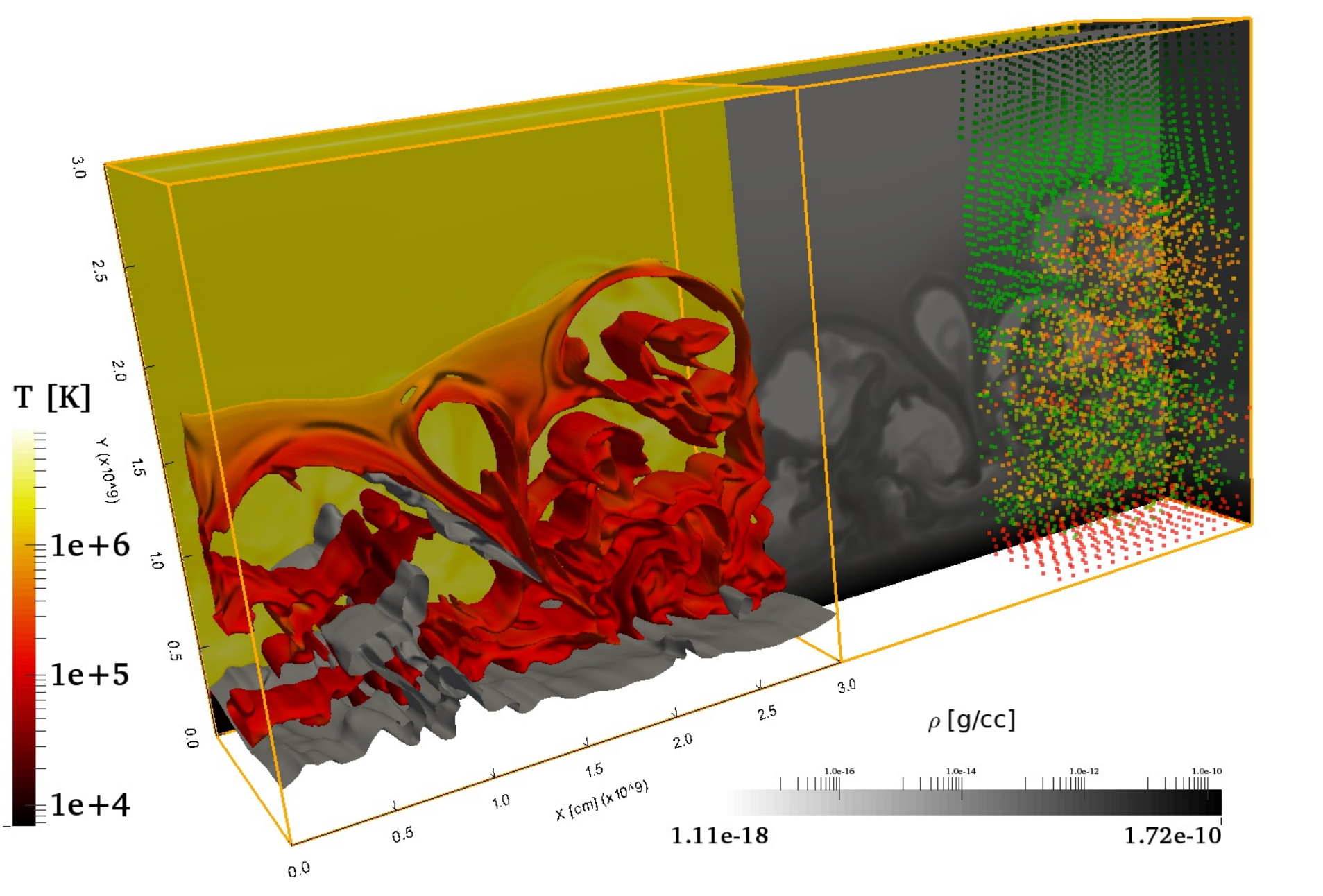}
\caption{After 14.3 minutes, this 3D view shows at left the temperature variation on the prominence boundary (in red to yellow), as well as (in grey) the location of the heavily perturbed chromosphere-corona transition region. The prominence is in a state of vigorous nonlinear magnetoconvection, also shown by its density variation in a cutting plane, and the tracer particles at right. The latter were originally arranged from green to red in plane-parallel fashion from top to bottom. While this figure is for the high field case, a movie in this format for the low field case is provided as online material.
}
\label{ffig4}
\end{figure}

\begin{figure}
\includegraphics[width=\textwidth]{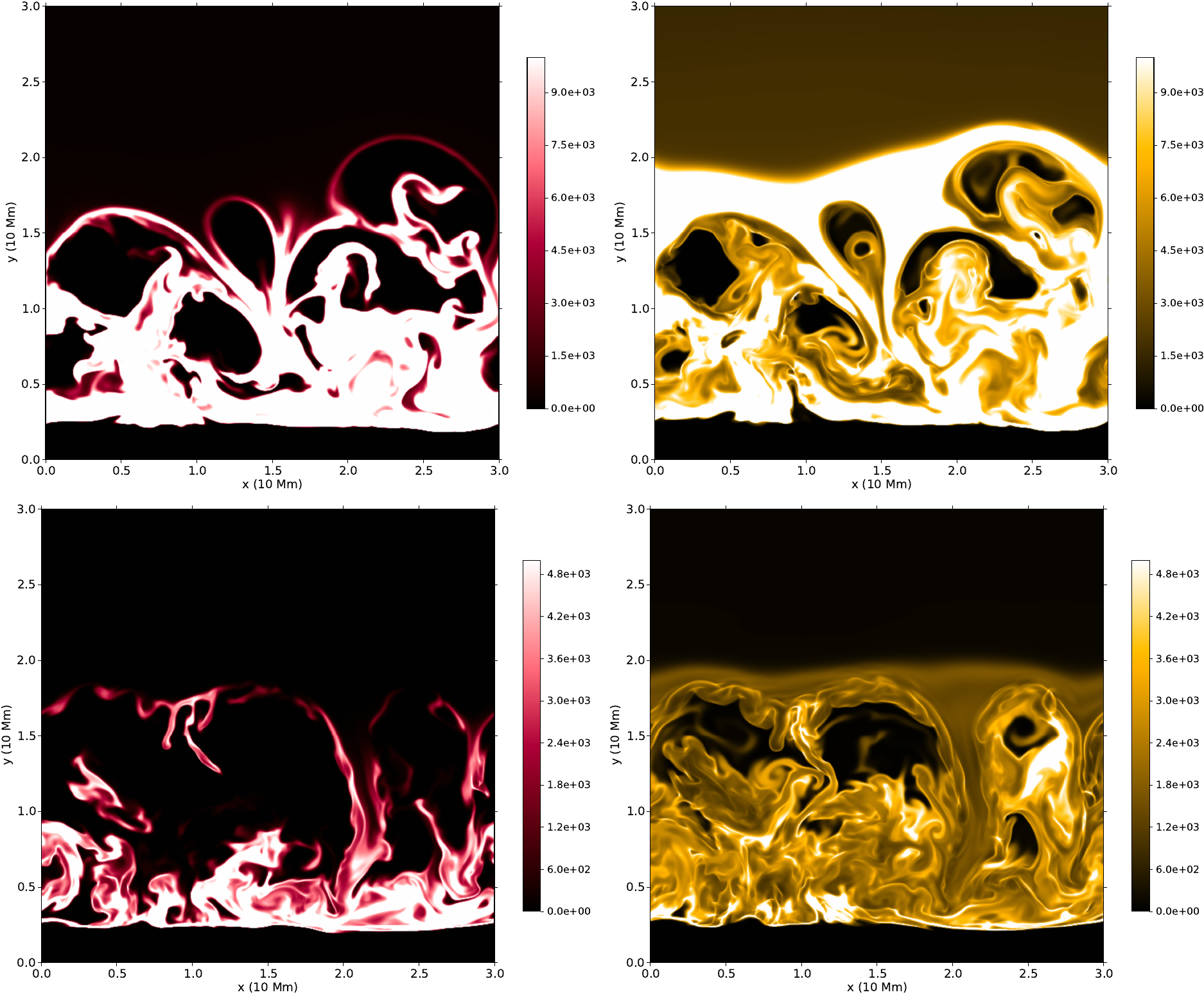}
\caption{SDO AIA views on the endstate after 14.3 minutes for both the high field case (top row) and low field case (bottom row). Left panels are at 304 \AA, right panels for 171 \AA. A movie comparing both cases in this format is given online, covering the last 7 minutes of evolution.
}
\label{ffig5}
\end{figure}

\end{document}